\documentclass[11pt]{article}
\usepackage[english]{babel}
\input{epsf}
\usepackage{mathtext}
\usepackage{graphics,dcolumn,bm,float}
\usepackage{amssymb,amsmath,rotate,color}
\usepackage{times}
\newcommand{\hhh}{{\cal H}}

\newcommand{\CE}{{\cal E}}
\newcommand{\be}{\begin{equation}}
\newcommand{\ene}{\end{equation}}
\newcommand{\ba}{\begin{array}}
\newcommand{\ea}{\end{array}}

\newcommand{\bsigma}{\mbox{\boldmath$\sigma$}}

\begin{document}

\title{Novel Majorana mode and magnetoresistance in ferromagnetic superconducting topological insulator}
\author{H. Goudarzi$^*$, M. Khezerlou$^{\dagger}$ and S. Asgarifar\\
\footnotesize\textit{Department of Physics, Faculty of Science, Urmia University, P.O.Box: 165, Urmia, Iran}}
\date{}
\maketitle

\begin{abstract}
Among the potential applications of topological insulators, we investigate theoretically the effect of coexistence of proximity-induced ferromagnetism and superconductivity on the surface states of 3-dimensional topological insulator, where the superconducting electron-hole excitations can be significantly affected by the magnetization of ferromagnetic order. We find that, Majorana mode energy, as a verified feature of TI F/S structure, along the interface sensitively depends on the magnitude of magnetization $m_{zfs}$ in FS region, while its slope in perpendicular incidence presents steep and no change. Since the superconducting gap is renormalized by a factor $\eta(m_{zfs})$, hence Andreev reflection is more or less suppressed, and, in particular, resulting subgap tunneling conductance is more sensitive to the magnitude of magnetizations in FS and F regions. Furthermore, an interesting scenario happens at the antiparallel configuration of magnetizations $m_{zf}$ and $m_{zfs}$ resulting in magnetoresistance in N/F/FS junction, which can be controlled and decreased by tuning the magnetization magnitude in FS region.
\end{abstract}

\section{INTRODUCTION}

Topological insulators (TIs) represent new type of material which has emerged in the last few years as one of the most actively research subjects in condensed matter physics. They are characterized by a full insulating gap in the bulk and gapless edge or surface states, which are protected by the time-reversal symmetry $\cite{1,1a,2,2a}$. Regarding Bernevig and Hughes prediction $\cite{2,3}$, TIs have been experimentally observed with such properties that host bound states on their surface, e.g. in 3-dimensional topological insulators (3DTI) $Bi_{2}Te_{3}$, $Bi_{2}Se_{3}$, $Sb_{2}Te_{3}$ and $Bi_{x}Sb_{1-x}$ alloy, and also in the $CdTe/HgTe/CdTe$ quantum well heterostructure $\cite{4,5,6}$. These states form a band-gap closing Dirac cone on each surface, and lead to a conducting state with properties unlike any other known electronic systems. In particular, conformity of the conduction and valence bands to each other in and around Dirac points in the first Brillouin zone, possessing an odd number of Dirac points, description of fermionic excitations as massless two-dimensional chiral Dirac fermions, depending chirality on the spin of electron, having the significant electron-phonon scattering on the surface, owning very low room-temperature electron mobility are the peculiar properties of electronic structure of TIs. Interestingly, the charge carriers in the surface states can behave as massive Dirac fermions $\cite{35}$ due to its proximity to a ferromagnetic material, that the vertical component of the magnetic vector potential may be proportional to the effective mass of Dirac fermion. The experimentally observed proximity-induced superconductivity on the surface state is another interesting dynamical feature occuring in 3DTI, see Refs. $\cite{12,13,15}$.

More importantly, the coexistence of superconductivity and ferromagnetism as one of potential interests for spintronics and high magnetic field applications has firstly been predicted by Fulde and Ferrel $\cite{17}$, and Larkin and Ovchinnikov $\cite{18}$ as \textit{FFLO state}. 
This effect can be in compliance with standard BCS theory for phonon-mediated $s$-wave superconductivity, because the ferromagnetic exchange field is expected to prevent spin-singlet Cooper pairing, (see, Ref. $\cite{29}$ as a prior work).
The magnetic polarization of a pair electron caused by a ferromagnetic material can lead to the different momentum of Cooper pair occurring in a ferromagnetic superconducting (FS) segment. It seems to be in contrast to the formation of a typical cooper pair, where two electrons may be in opposite spin direction with the same momentum.
However, Bergeret et.al. $\cite{19}$ and Li et.al. $\cite{20}$ have studied the effect of superconductor/ferromagnetic bilayer on the critical Josephson current, where the orientation of ferromagnetic exchange field strongly affects the critical current. Also, the effect of superconductivity in coexistence with ferromagnetism has been studied on the superconducting gap equation for two case of singlet $s$-wave and triplet $p$-wave symmetries $\cite{21}$. The authors have reconsidered the Clogston-Chandrasekhar limiting $\cite{CC,CH}$. According to the Clogston criterion in the conventional FS mixture, the normal state is regained as soon as the ferromagnetic exchange field exceeds $\Delta_0/\sqrt{2}$ at zero temperature.
To be empirically, the $ErRh_{4}B_{4}$ $\cite{22}$ has been discovered to be the first ferromagnetic superconductor, which superconductivity is found to occur in a small temperature interval with adjusted ferromagnetic phase. Also, superconductivity is detected in itinerant ferromagnetic $UGe_{2}$ in a limited range of pressure and temperature $\cite{23}$. 

Regarding several works in the recent few years concerning with the topological insulator-based junctions $\cite{7,8,39,40,46,48,j1,j2,j3,j4,j5,j6,j8,j7}$, which are related to the Andreev process and resulting subgap conductance we proceed, in this paper, to theoretically study the dynamical properties of Dirac-like charge carriers in the surface states of 3DTI under influence of both superconducting and ferromagnetic orders via the introducing the proper form of corresponding Dirac spinors, which are principally distinct from those given in Ref. $\cite{j1}$.
The magnetization induction opens a gap at the Dirac point (no inducing any finite center of mass momentum to the Cooper pair), whereas the superconducting correlations causes an energy gap at the Fermi level in the 3DTI. It will be particularly interesting to investigate the topological insulator superconducting electron-hole excitations in the presence of a exchange field. We assume that the Fermi level is close to the Dirac point, and the ferromagnet has a magnetization $|\textbf{M}|<\mu$. The chirality conservation of charge carriers on the surface states in the presence of magnetization (due to opening the band gap) allows to use a finite magnitude of $|\textbf{M}|$. In the absence of topological insulator, the spin-splitting caused by magnetization gives rise to limiting the magnitude of $|\textbf{M}|$ in a FS structure.
These excitations, therefore, are found to play a crucial role in Andreev reflection (AR) process leading to the tunneling conductance below the renormalized superconducting gap. Particularly, we pay attention to the formation of Majorana bound energy mode, as an interesting feature in topological insulator ferromagnet/superconductor interface, depending on the magnetization of FS hybrid structure. We present, in section 2, the explicit signature of magnetization in low-energy effective Dirac-Bogogliubov-de Gennes (DBdG) Hamiltonian. The electron(hole) quasiparticle dispersion energy is analytically calculated, which seems to exhibit qualitatively distinct behavior in hole excitations ($|k_{fs}|<k_F$) by varying the magnitude of magnetization. By considering the magnetization is ever less than chemical potential in FS region, the superconducting wavevector and corresponding eigenstates are derived analytically. Section 3 is devoted to unveil the above key point of FS energy excitation, Majorana mode energy, Andreev process and resulting tunneling conductance in N/F/FS junction and respective discussions. In the last section, the main characteristics of proposed structure are summarized. 

\section{THEORETICAL FORMALISM}
\subsection{Topological insulator FS effective Hamiltonian}

In order to investigate how both superconductivity and ferromagnetism induction to the surface state affects the electron-hole excitations in a 3DTI hybrid structure, we consider magnetization contribution to the DBdG equation.
Let us focus first on the Hubbard model Hamiltonian $\cite{HU}$ that is included the effective exchange field $\mathbf{M}$ follows from:
\begin{equation}
\hhh=-\sum_{\rho\rho^{'}s}t_{\rho\rho^{'}}\hat{c}^{\dagger}_{\rho s}\hat{c}_{\rho^{'}s}+\frac{1}{2}\sum_{\rho\rho^{'}ss^{'}}U_{\rho\rho^{'}ss^{'}}\hat{n}_{\rho s}\hat{n}_{\rho^{'}s^{'}}+\sum_{\rho ss^{'}}\hat{c}^{\dagger}_{\rho s}(\bsigma\cdot\mathbf{M})\hat{c}_{\rho s^{'}},
\end{equation}
where $U_{\rho\rho^{'}ss^{'}}$ denotes the effective attractive interaction between arbitrary electrons, labeled by the integer $\rho$ and $\rho^{'}$ with spins $s$ and $s^{'}$. The matrices $t_{\rho\rho^{'}}$ are responsible for the hopping between different neighboring sites, and $\hat{c}_{\rho s}$ and $\hat{n}_{\rho s}$ indicate the second quantized fermion and number operators, respectively. Here, $\bsigma (\sigma_x,\sigma_y,\sigma_z)$ is the vector of Pauli matrix. Using the Hartree-Fock-Gorkov approximation and Bogoliubov-Valatin transformation $\cite{HB}$, the Bogoliubov-de Gennes Hamiltonian describing dynamics of Bogoliubov quasiparticles is found. In Nambu basis, that electron(hole) state is given by $\Psi=\left(\psi_{\uparrow},\psi_{\downarrow},\psi^{\dagger}_{\uparrow},\psi^{\dagger}_{\downarrow}\right)$, the BdG Hamiltonian for a $s$-wave spin singlet superconducting gap in the presence of an exchange splitting can be written as:  
\begin{equation}
\hhh_{SF}=\left(\begin{array}{cc}
h(\mathbf{k})+M&\Delta(\mathbf{k})\\
-\Delta^{\ast}(-\mathbf{k})&-h^{\ast}(-\mathbf{k})-M
\end{array}\right),
\label{2}
\end{equation}
where $h(\mathbf{k})$ denotes the non-superconducting Schrodinger-type part, and $\Delta(\mathbf{k})$ is superconducting order parameter. In the simplest model, $\Delta(\mathbf{k})$ can be chosen to be real to describe time-reversed states. The effective exchange field by rotating our spin reference frame can be gain as $\left|\mathbf{M}\right|=\sqrt{m^{2}_{x}+m^{2}_{y}+m^{2}_{z}}$. The four corresponding levels of a singlet superconductor in a spin magnetic field is obtained $E_{s}(\mathbf{k})=\sqrt{\epsilon^{2}_{\mathbf{k}}+\left|\Delta(\mathbf{k})\right|^{2}}+s \left|\mathbf{M}\right|$ with $s=\pm 1$, where $\epsilon_{\mathbf{k}}$ is the normal state energy for $h(\mathbf{k})$. However, dependence of superconducting order parameter on the exchange energy can be exactly derived from self-consistency condition $\cite{21}$:
\begin{equation}
\Delta(\mathbf{k})=-\frac{1}{4}\sum_{\mathbf{k}s}U_{s-s}(\mathbf{k})\frac{\Delta_0(\mathbf{k})}{\sqrt{\epsilon^{2}_{\mathbf{k}}+\left|\Delta_0(\mathbf{k})\right|^{2}}}\tanh\left(\frac{\sqrt{\epsilon^{2}_{\mathbf{k}}+\left|\Delta_0(\mathbf{k})\right|^{2}}+s \left|\mathbf{M}\right|}{2k_BT}\right),
\label{a}
\end{equation}
where $\Delta_0(\mathbf{k})$ is the conventional order parameter in absence of ferromagnetic effect, $k_B$ and $T$ are the Boltzmann constant and temperature, respectively. The exchange splitting dependence of superconducting gap indicates that equation \eqref{a} has no functionality of $\mathbf{M}$ at zero temperature. This takes place under an important condition known as Clogston-Chandrasekhar limiting $\cite{CC,CH}$. According to this condition, if the exchange splitting becomes greater than a critical value $\left|\mathbf{M}_c\right|=\left|\Delta(T=0)\right|/\sqrt{2}$, then the normal state has a lower energy than the superconducting state. This means that a  phase transition from the superconducting to normal states is possible when the exchange splitting is increased at zero temperature.

We now proceed to treat such a ferromagnetic superconductivity coexistence at the Dirac point of a 3DTI. It should be stressed that the dressed Dirac fermions with an exchange field in topologically conserved surface state have to be in superconducting state. Here, the influence of exchange field interacts in a fundamentally different way comparing to the conventional topologically trivial system, where the exchange field splits the energy bands of the majority and minority spins. A strong TI is a material that the conducting surface states at an odd number of Dirac points in the Brillouin zone close the insulating bulk gap unless time-reversal symmetry is broken. Candidate Dirac-type materials include the semiconducting alloy $Bi_{1-x}Sb_x$, as well as HgTe and $\alpha-Sn$ under uniaxial strain $\cite{STI}$. In the simplest case, there is a single Dirac point in the surface Fermi circle and general effective Hamiltonian is modeled as $h^{TI}_{N}=\hbar v_{F}(\bsigma\cdot\mathbf{k})-\mu$,
where $v_{F}$ indicates the surface Fermi velocity, and $\mu$ is the chemical potential. Under the influence of a ferromagnetic proximity effect, the Hamiltonian for the two-dimensional surface states of a 3DTI reads as:
$$
h^{TI}_{F}=\hbar v_{F}(\bsigma\cdot\mathbf{k})-\mu+\mathbf{M}\cdot\bsigma,
$$
where the ferromagnetic contribution corresponds to an exchange field $\mathbf{M}=(m_{x},m_{y},m_{z})$. It has been shown $\cite{35}$ that transverse components of the magnetization on the surface $(m_{x},m_{y})$ are responsible to shift the position of the Fermi surface of band dispersion, while its perpendicular component to the surface induces an energy gap between conduction and valence bands.

In what follows, we will employ the relativistic generalization of BdG Hamiltonian, which is interacted by the effective exchange field to obtain the dispersion relation of FS dressed Dirac electrons in a topological insulator:
\begin{equation}
\hhh^{TI}_{FS}=\left(\begin{array}{cc}
h^{TI}_{F}(\mathbf{k})&\Delta(\mathbf{k})\\
-\Delta^{\ast}(-\mathbf{k})&-h^{TI\ast}_{F}(-\mathbf{k})
\end{array}\right).
\label{3}
\end{equation}
The superconducting order parameter now depends on both spin and momentum symmetry of the Cooper pair, that the gap matrix for spin-singlet can be given as $\Delta(\mathbf{k})=i\Delta_{0}\sigma_{y}e^{i\varphi}$,
where $\Delta_{0}$ is the uniform amplitude of the superconducting gap and phase $\varphi$ guarantees the globally broken $U(1)$ symmetry. By diagonalizing this Hamiltonian we arrive at an energy-momentum quartic equation. Without lose of essential physics, we suppose the component of magnetization vector along the transport direction to be zero $m_{x}=0$ for simplicity. Also, we set $m_{y}=0$, since the analytical calculations become unwieldy otherwise. The dispersion relation resulted from Eq. \eqref{3} for electron-hole excitations  is found to be of the form:
\begin{equation}
\CE_{FS}=\zeta\sqrt{\left(-\tau\mu_{fs}+\sqrt{m^{2}_{zfs}+\left|\mathbf{k}_{FS}\right|^{2}+\left|\Delta_0\right|^{2}(\frac{m_{zfs}}{\mu_{fs}})^{2}}\right)^{2}+\left|\Delta_0\right|^{2}\left(1-(\frac{m_{zfs}}{\mu_{fs}})^{2}\right)},
\label{5}
\end{equation}
where, the parameter $\zeta=\pm 1$ denotes the electron-like and hole-like excitations, while $\tau=\pm 1$ distinguishes the conduction and valence bands. We might expect several anomalous properties from the above superconducting excitations, which is investigated in detail in the next section. Equation \eqref{5} is clearly reduced to the standard eigenvalues for superconductor topological insulator in the absence of exchange field as $m_{z}=0$ (see Ref. $\cite{35}$), $\CE_{S}=\zeta\sqrt{\left(-\tau\mu_s+\left|\mathbf{k}_{S}\right|\right)^2+\left|\Delta_0\right|^{2}}$. The mean-field conditions are satisfied as long as $\Delta_{0}\ll\mu_{fs}$. In this condition, the exact form of superconducting wavevector of charge carriers can be acquired from the eigenstates $k_{fs}=\sqrt{\mu^2_{fs}-m^2_{zfs}}$.

The Hamiltonian Eq. \eqref{3} can be solved to obtain the electron (hole) eigenstates for FS topological insulator. The wavefunctions including a contribution of both electron-like and hole-like quasiparticles are analytically found as:
\begin{equation}
\psi^{e}_{FS}=\left(\begin{array}{cc}
e^{i\beta}\\
e^{i\beta}e^{i\theta_{fs}}\\
-e^{i\theta_{fs}}e^{-i\gamma_{e}}e^{-i\varphi}\\
e^{-i\gamma_{e}}e^{-i\varphi}
\end{array}\right)e^{i(k_{fs}^{x}x+k_{fs}^{y}y)}, \ \ \psi^{h}_{FS}=\left(\begin{array}{cc}
1\\
-e^{-i\theta_{fs}}\\
e^{i\beta}e^{-i\theta_{fs}}e^{-i\gamma_{h}}e^{-i\varphi}\\
e^{i\beta}e^{-i\gamma_{h}}e^{-i\varphi}
\end{array}\right)e^{i(-k_{fs}^{x}x+k_{fs}^{y}y)},
\end{equation}
where we define 
$$
\cos{\beta}=\frac{\CE_{FS}}{\eta\left|\Delta_0\right|} \ ; \ \ \eta=\sqrt{1-(\frac{m_{zfs}}{\mu_{fs}})^{2}} \ , \ \ e^{i\gamma_{e(h)}}=\frac{\Delta(\mathbf{k})}{\left|\Delta(\mathbf{k})\right|}.
$$
Note that, the solution is allowed as long as the Zeeman field being lower than chemical potential $m_{zfs}\leq \mu_{fs}$.

\subsection{FS interplay at the TI interface}

We consider Andreev reflection in a hybrid N/F/FS structure formed on the surface of a 3DTI which coexistence between ferromagnet and superconductor is assumed to be induced by means of the proximity effect. The wide topological insulator junction is taken along the $x$-axis with the FS region for $x>L$, F region for $0<x<L$ and N region for $x<0$. The superconducting order parameter vanishes identically in N and F regions, and we can neglect its spatial variation in the FS region close to the interface. The magnetization vectors of both sections is taken, in general, $m_{zi} (i\equiv f,fs)$, which can be at the parallel or antiparallel configuration, as shown in Fig. 1.
In the scattering process follows from the Blonder-Tinkham-Klapwijk (BTK) formula $\cite{BTK}$, we find the reflection amplitudes from the boundary condition at the interface.
In ferromagnetic case, right- and left- moving electrons (holes) with energy excitation $\epsilon_F=\pm\sqrt{k_{Ff}^{2}+m_{zf}^2}-\mu_{f}$ below the superconducting gap, transmitted (normal reflected) from the N region and reflected (Andreev reflected) at the FS interface. Thus, the leftover $2e$ charge is transferred into the FS region as a Cooper pair at Fermi level. At energy excitation above the normalized superconducting gap resulted from Eq. (5) (see, in particular, Fig. 2) quasiparticle states can directly tunnel into the superconducting section. The reflected hole leading to AR can be actually controlled by the doping level in order to take place possible specular Andreev reflection. Particularly, we have to determine (via the dynamical features of system) the allowed values of Fermi energy in three regions. We set the Fermi energy to zero in F region. The electron(or hole) transmitted to the FS region angle may be accordingly obtained from the fact of conservation of transverse wavevector under quasiparticle scattering at the interface: 
\begin{equation}
\theta_{fs}=\arcsin\left(\frac{\mu_n\sin{\theta}}{\sqrt{\mu^{2}_{fs}-m^{2}_{zfs}}}\right),
\label{7}
\end{equation}
where $\mu_n$ and $\theta$ are the chemical potential and incidence angle in N region, respectively. As an important point, the electron(hole) angle of incidence in all regions may be span the range from $0$ to $\pi/2$ around the normal axis. Regarding the Eqs. \eqref{7}, the angle $\theta_{fs}$ needs to be meaningful when the chemical potential of FS region takes a magnitude greater than its value in N region ($\mu_{fs}>\mu_{n}$). On the other hand, we previously applied the condition $m_{zfs}<\mu_{fs}$, as an experimentally used manner to calculate the wave functions Eq. (6). 

By introducing the normal $r$ and Andreev $r_A$ reflection coefficients and the scattering coefficients in F region, the total wave function inside the N and F region can be written as:
$$
\Psi_{N}=e^{ik^{y}_{n}y}\left(\psi^{e+}_{N}e^{ik^{x}_{n}x}+r\psi^{e-}_{N}e^{-i k^{x}_{n}x}+r_{A}\psi^{h-}_{N}e^{i k^{x}_{n}x}\right),
$$
\begin{equation}
\Psi_{F}=e^{ik^{y}_{f}y}\left(a\psi^{e+}_Fe^{ik^{xe}_{f}x}+b\psi^{e-}_Fe^{-ik^{xe}_{f}x}+c\psi^{h+}_Fe^{-i k^{xh}_{f}x}+d\psi^{h-}_Fe^{i k^{xh}_{f}x}\right),
\label{6}
\end{equation}
where the eigenvectors $\psi$ can be found in Appendix A. The probability amplitude of reflections in Eq. \eqref{6} are calculated from the continuity of the wavefunctions at the interface. The wave function in FS region is defined as $\Psi_{FS}=t^e\psi^{e}_{FS}+t^h\psi^{h}_{FS}$. Finally, we find the following analytical expressions for the reflection coefficients, that the auxiliary quantities is described in Appendix A:
$$
r=\left[t^ee^{i\beta}(2\mathcal{M}_1-1)+t^he^{-i\beta}(2\mathcal{M}_2-1)\right](i\sin{(k^{xe}_{f}L}))+
$$
$$
\left[t^ee^{i\beta}+t^he^{-i\beta}\right]\cos{(k^{xe}_{f}L})-1,
$$
$$
r_A=\left[t^ee^{i\theta_{fs}}(2\mathcal{M}_2-1)-t^he^{-i\theta_{fs}}(2\mathcal{M}_1-1)\right](i\sin{(k^{xe}_{f}L}))-
$$   
\begin{equation}
\left[t^ee^{i\theta_{fs}}-t^he^{-i\theta_{fs}}\right]\cos{(k^{xe}_{f}L)}.
\label{r}
\end{equation}
The reflection amplitudes measurements under the BTK formalism enables us to capture the tunneling conductance through the junction:
\begin{equation}
G(eV)=G_{0}\int^{\theta_c}_{0}d\theta^{e}\cos{\theta^e}\left(1+\left|r_A\right|^2-\left|r\right|^2\right),
\label{9}
\end{equation}
where the critical angle of incidence $\theta_c$ is determined depending on the doping of F region. The quantity $G_0$ is a renormalization factor corresponding to the ballistic conductance of normal metallic junction.

\section{RESULTS AND DISCUSSION}
\subsection{Energy excitation and Majorana mode}

In this section, we proceed to analyze in detail the dynamical features of Dirac-like charge carriers in 3DTI with ferromagnetic and superconducting orders deposited on top of it. We assume that the Fermi level controlled by the chemical potential $\mu$ is close to the Dirac point. In this case, it is expected the signature of $m_{zfs}<\mu_{fs}$ to be significant. In Fig. 2, we demonstrate the FS 3DTI electron-hole excitations. A net superconducting gap $\Delta_0$ is obtained in Dirac points (for $|k_{fs}|=k_F$, where $k_F$ is Fermi wavevector) when we set $m_{zfs}=0$. Increasing $m_{zfs}$ up to its possible maximum value results in three outcomes: i) the superconducting excitations, which is renormalized by a factor $|\Delta(\textbf{k})|\sqrt{1-(m_{zfs}/\mu_{fs})^2}$, disappear in hole branch ($|k_{fs}|<k_F$). It means that for the greater magnetizations, if we consent the superconductivity in FS 3DTI still exists, there is almost vanishing quantum state for reflected hole by Andreev process in the valence band, ii) Dirac point is shifted towards smaller FS quasiparticle electron-hole wavevectors, iii) the superconducting gap decreases slowly, where the variation of net gap is very low $\delta\Delta_0\ll|\Delta(\textbf{k})|$. The Andreev process, therefore, is believed to inconsiderably supress. The signature of these valence band excitations can be clearly shown in AR, where the Majorana mode may also be formed at the 3DTI F/FS interface $\cite{7,39}$.

As a verified result, considering the topological insulator interface between the ferromagnetic insulator and conventional superconductor leads to the appearance of the chiral Majorana mode as an Andreev bound state. In other words, the Majorana mode and Andreev reflection are strongly related to each other. The latter can be realized by the fact of looking for bound energies produced by the perfect AR, which yields the following solution:
$$
\tilde{\epsilon}(\theta)=\eta\Delta_0 sgn\left(\Lambda\right)/\sqrt{1+\Lambda^2} \ ; \ \ \ \Lambda=\tan\left[\frac{1}{2i}\ln(\frac{\upsilon_1}{\upsilon_2})\right],
$$
where we define
$$
\upsilon_{1(2)}=4i\sin{k^{xe}_{f}L}\cos{\theta}\mathcal{M}_{2(1)}\mathcal{A}_{1(2)}+2e^{-ik^{xe}_{f}L}\cos{\theta}\mathcal{A}_{1(2)}-\mathcal{B}_{2(1)}\mathcal{A}_{1(2)}.
$$
We have checked numerically that sign of $\Lambda$ is changed by $sgn(m_{zf})$. Thus, the sign of Andreev resonance states may be changed by reversing the direction of $m_{zf}$, and it corresponds to the chirality of Majorana mode energies. As shown in Fig. 3, the slope of the energy curves of $\tilde{\epsilon}(\theta)$ around $\tilde{\epsilon}(\theta=0)=0$ become steep and show no change with the increase of $m_{zfs}/\mu_n$ for fixed $m_{zf}$, while it exhibits significantly decreasing behavior with the increase of $m_{zf}/\mu_n$ for fixed $m_{zfs}$. The dispersion of Majorana modes along the interface ($\theta=\pi/2$) decreases with the increase of both magnetizations of FS and F regions. Note that, due to the presence of $m_{zfs}$ it needs to consider the Fermi level mismatch between normal and FS sections, i.e. $\mu_n\neq\mu_{fs}$. Then, the above contributions can be considerable in Andreev process and resulting subgap tunneling conductance.

\subsection{Tunneling conductance}

From the angle-resolved Andreev and normal reflection probabilities using Eq. \eqref{r}, we see from Fig. 4(a) the main contribution of AR belongs to the angle of incidence $\theta<0.15\pi$ in zero bias. It, therefore, is expected to achieve the lower zero bias conductance, as shown in Fig. 4(b) and (c). Furthermore, varying $m_{zfs}$ has no significant influence on AR in zero bias $\epsilon(eV)=0$ owing to the very small decrease of the renormalized superconducting gap with the increase of the $m_{zfs}$, while the increasing $m_{zf}$ results in more suppression of AR. The latter can be understood by the increase of band gap in Dirac point in F region. The resulting normalized angle-averaged tunneling conductance curves are reported in Figs. 4(b) and (c) for two parallel and antiparallel configurations of magnetizations in FS and F regions. Zero bias conductance peak disappears with the decrease of the $m_{zf}$, and instead of it a high conductance peak appears in bias $\epsilon =\eta\Delta_0$. This result should be compared to that is obtained in Ref. $\cite{35,8}$. Interestingly, by increasing the $m_{zfs}$ the magnitude of subgap bias $\epsilon/\eta\Delta_0<1$, for which the new peak takes place, is limited, as seen from Fig. 4(c). Thus, parameter $\eta=\sqrt{1-(m_{zfs}/\mu_s)^2}$ can be considered as a ``\textit{bias-limitation coefficient}". These features have been obtained when the direction of magnetizations in F and FS regions are at the parallel configuration. The fundamentally distinct scenarios we find for the case of antiparallel configuration of magnetizations. In this case, first, the tunneling subgap conductance is enhanced, secondly, the zero bias conductance peak presented in parallel case is replaced by a deep, see, in detail, Fig. 4(d). Dynamically description, when the direction of $m_{zf}$ is inverted we, indeed, meet with an inverse energy gap in Dirac point of 3DTI giving rise to enhancing the conductance peak respective bias energy $\epsilon/\Delta_0=\eta$ in low values of $m_{zf}$. One can express that the zero bias conductance originates from the chiral Majorana mode, which significantly depends on the $m_{zf}$. The chirality actually corresponds to the sign of $m_{zf}$, while the magnitudes of zero bias conductance at the parallel and antiparallel configurations are the same. Hence, the both deep and peak of conductance curves in antiparallel case are significance being influenced by the inverted gap caused by the $-m_{zf}$.

Remarkably, the importance of above findings can be featured by the capture of magnetoresistance (MR) of the topological insulator junction. The magnetization (specialy in FS region) dependence of MR is presented in Fig. 5, where we observe a considerable MR peak for extra values of $m_{zf}$ (e.g. $0.9\mu_n$ in figure). Importantly, increasing the $m_{zfs}$ weakens the MR peak, since, regarding the superconducting excitations in Fig. 2, the AR is more or less suppressed in the presence of $m_{zfs}$ and Fermi wavevector mismatch also causes to decrease the $\eta\Delta_0$-bias conductance peak at the antiparallel configuration. According to the conductance curves, there is no MR in zero bias.

\section{CONCLUSION}

In summary, we have investigated the influence of ferromagnetic superconducting orders coexistence in the surface state of topological insulator. The topological insulator superconducting electron-hole excitations in the presence of magnetization have led to achieve qualitatively distinct transport properties in tunneling N/F/FS junction. One of key findings of the present work is that the resulting subgap conductance has been found to be strongly sensitive to the parallel or antiparallel configuration of magnetization directions in FS and F regions. Thus, this feature has actually led to present the magnetoresistance peak for bias energy close to the renormalized superconducting gap $\epsilon(eV)=\eta\Delta_0$, which the bias limitation coefficient $\eta$ includes the magnetization of FS region $m_{zfs}$. Particularly, we have found the presence of Majorana mode at the F/FS interface to be controlled by the tuning of magnetizations magnitude. However, these results have been obtained in the case of $m_{zfs}, m_{zf}<\mu$ and $\mu\gg\Delta_0$, which is relevant to the experimental regime.

\renewcommand{\theequation}{A-\arabic{equation}}
  \setcounter{equation}{0}  
  \section*{APPENDIX A: Normal and Andreev reflection amplitudes}  

To complete calculation of probability of reflections in N/F/FS junction, we write down right and left moving electron and hole spinors in F and N region:
$$
\psi^{e+}_{N}=\left[1,e^{i\theta},0,0\right]^{T},\ \ \psi^{e-}_{N}=\left[1,-e^{-i\theta},0,0\right]^{T},\ \ \psi^{h-}_{N}=\left[0,0,1,-e^{-i\theta}\right]^{T},
$$
$$
\psi^{e+}_{F}=\left[1,\alpha e^{i\theta_f},0,0\right]^{T},\ \psi^{e-}_{F}=\left[1,-\alpha e^{-i\theta_f},0,0\right]^{T}, \ \psi^{h+}_{F}=\left[0,0,1,\alpha e^{i\theta_f}\right]^{T}, \ \psi^{h-}_{F}=\left[0,0,1,-\alpha e^{-i\theta_f}\right]^{T}
$$
where we define $\alpha=\sqrt{\frac{\mu_{f}-m_{zf}}{\mu_{f}+m_{zf}}}$. By matching boundary conditions on $\Psi_{N}$ and $\Psi_{F}$ at $x=0$ and $\Psi_{F}$ and $\Psi_{FS}$ at $x=L$, the reflection amplitudes are obtained. We introduce auxiliary quantities in Eq. \eqref{r} as:
$$
t^e=\frac{2\mathcal{A}_2\cos{\theta}}{\mathcal{B}_1\mathcal{A}_2e^{i\beta}-\mathcal{B}_2\mathcal{A}_1e^{-i\beta}},\ \ \ t^h=\frac{-2\mathcal{A}_1\cos{\theta}}{\mathcal{B}_1\mathcal{A}_2e^{i\beta}-\mathcal{B}_2\mathcal{A}_1e^{-i\beta}}.
$$
with
$$
\mathcal{A}_1(2)=e^{(-)i\theta_{fs}}\left[(-)\mathcal{N}_1(\mathcal{M}_{2(1)}-1)e^{ik^{xe}_{f}L}-(+)\mathcal{N}_2\mathcal{M}_{2(1)}e^{-ik^{xe}_{f}L}\right],
$$
$$
\mathcal{B}_1(2)=-\mathcal{N}_1(\mathcal{M}_{1(2)}-1)e^{-ik^{xe}_{f}L}+\mathcal{N}_2\mathcal{M}_{1(2)}e^{ik^{xe}_{f}L},
$$
$$
\mathcal{N}_1(2)=(-)\alpha e^{(-)i\theta_f}+e^{-i\theta},\ \ \ \mathcal{M}_{1(2)}=\frac{\alpha e^{i\theta_f}-(+)e^{(-)i\theta_{fs}}}{2\alpha\cos{\theta_f}}.
$$

$^*$h.goudarzi@urmia.ac.ir ; goudarzia@phys.msu.ru\\
$^{\dagger}$m.khezerlou@urmia.ac.ir\\

\newpage

\textbf{Figure captions}\\
\textbf{Figure 1} (color online) Sketch of the topological insulator-based N/F/FS junction. The magnetization vectors in F and FS regions can be at the parallel or antiparallel configuration.\\
\textbf{Figure 2} (color online) The ferromagnetic superconducting excitation spectra on the surface state of 3DTI for several values of $m_{zs}$, calculated from Eq. (5). We set the net value of superconducting gap $|\Delta_S|=0.5\;eV$ (this value of pair potential is taken only to more clarify the behavior of spectra in Dirac point, although it does not further need to use it in our calculations, since $\mu_{fs}/|\Delta_S|\gg 1$ is supposed.\\
\textbf{Figure 3} (color online) The dispersion of Majorana modes as a function of the electron incident angle for several values of magnetizations in FS and F regions. The solid lines correspond to $m_{zf}=0.2\mu_n$ and the dashed lines to $m_{zfs}=0.2\mu_n$.\\
\textbf{Figure 4(a), (b), (c), (d)} (color online) (a) Probability of the normal and Andreev reflections as a function of electron incidence angle at the interface in zero bias $\epsilon(eV)/\eta\Delta_0=0$ with $m_{zfs}=0.5\mu_n$ and $\mu_{fs}/\mu_n=1.5$. The plots show the results for different values of $m_{zf}$. (b) Normalized tunneling conductance versus bias voltage $eV$ and magnetization of F region. We set $m_{zfs}=0.5\mu_n$ (c) Normalized tunneling conductance versus bias voltage and magnetization of FS region. We set $m_{zf}=0.2\mu_n$ (d) The tunneling conductance as a function of bias voltage for two $\pm$ signs of $m_{zf}$, corresponding to the parallel and antiparallel configurations in F and FS regions. The solid lines correspond to $+m_{zf}$ and marker dashed lines correspond to $-m_{zf}$. We set $m_{zf}=0.2\mu_n$.\\
\textbf{Figure 5} (color online) The magnetoresistance spectra as function of bias voltage, where the influence of $m_{zfs}$ and $m_{zf}$ is indicated, separately. We have set $\mu_{fs}/\mu_n=1.2$ in the resulting conductance and magnetoresistance spectra.

\newpage

\begin{figure}[p]
\epsfxsize=0.4 \textwidth
\begin{center}
\epsfbox{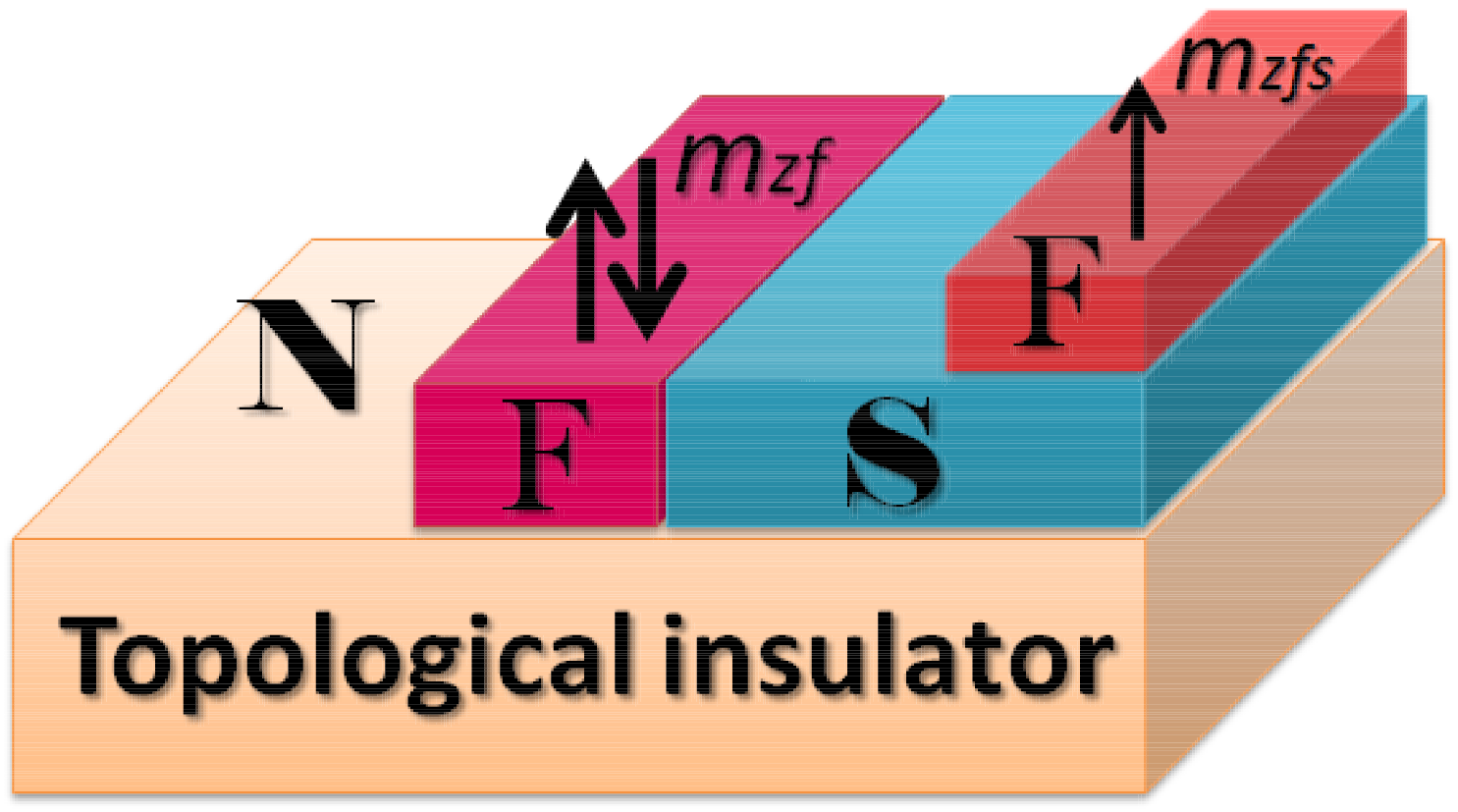}
\setcounter{figure}{0}
\caption{\footnotesize }
\end{center}
\end{figure}

\begin{figure}[p]
\epsfxsize=0.5 \textwidth
\begin{center}
\epsfbox{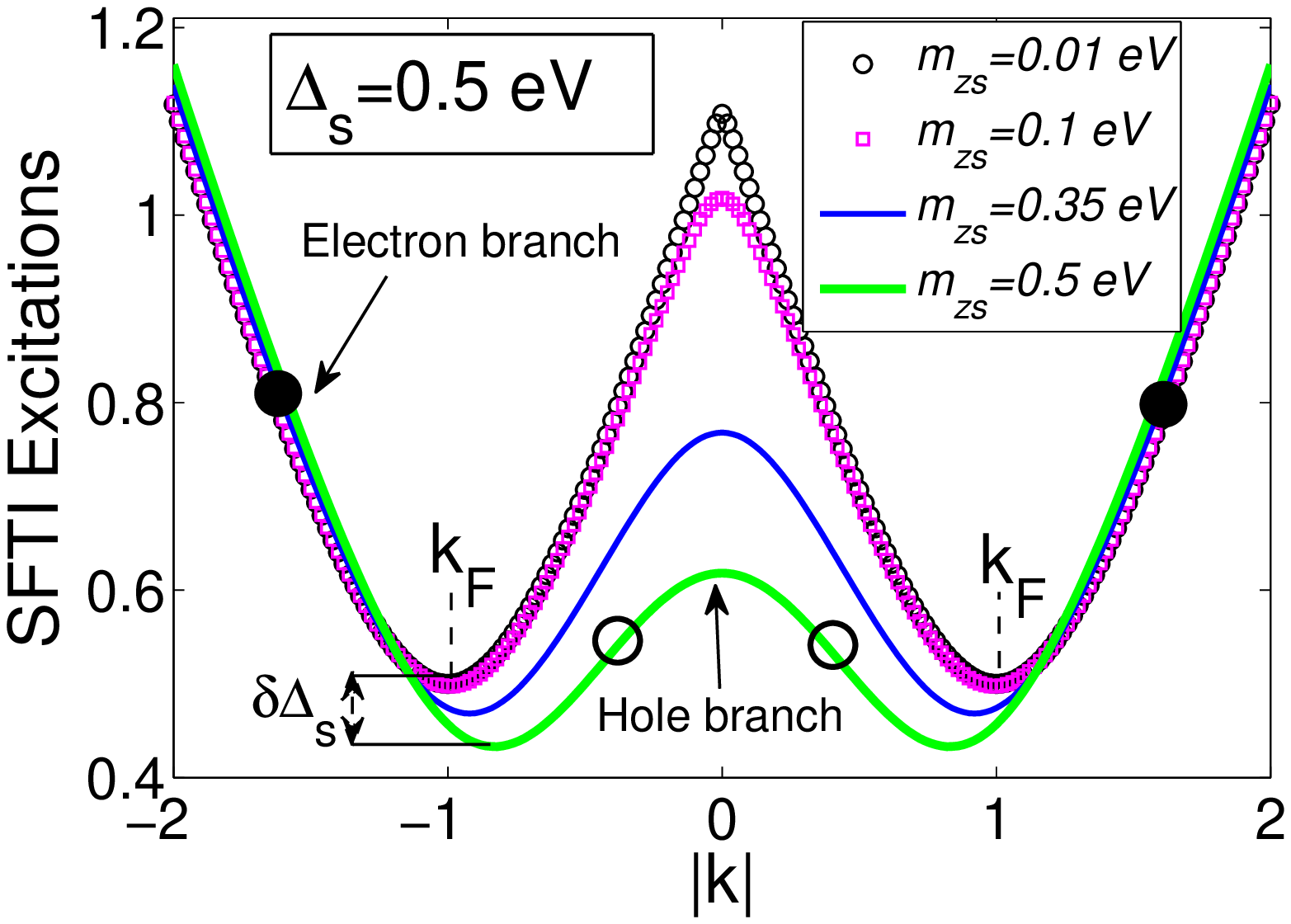}
\setcounter{figure}{1}
\caption{\footnotesize }
\end{center}
\end{figure}

\begin{figure}[p]
\epsfxsize=0.5 \textwidth
\begin{center}
\epsfbox{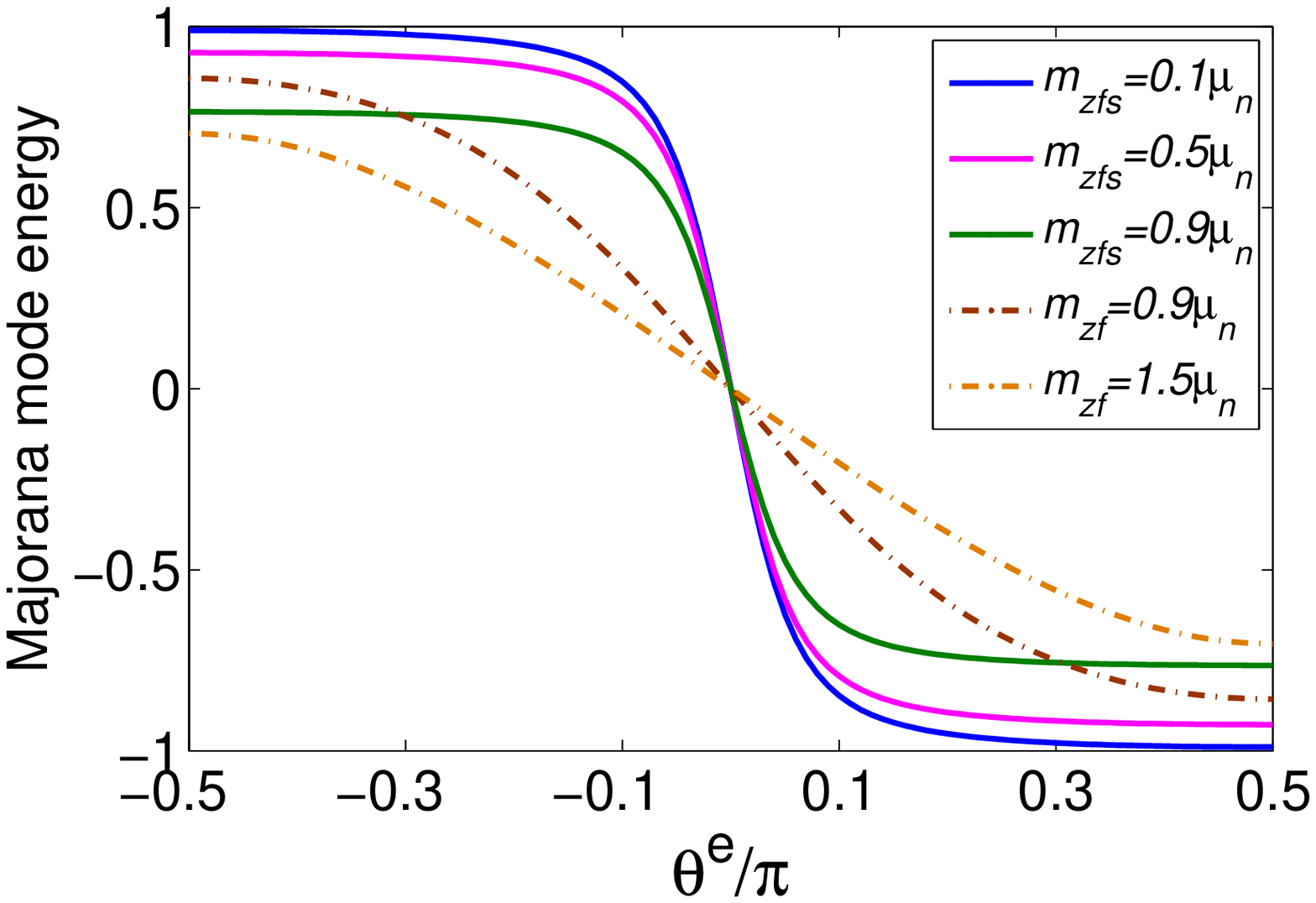}
\setcounter{figure}{2}
\caption{\footnotesize }
\end{center}
\end{figure}

\begin{figure}[p]
\epsfxsize=0.5 \textwidth
\begin{center}
\epsfbox{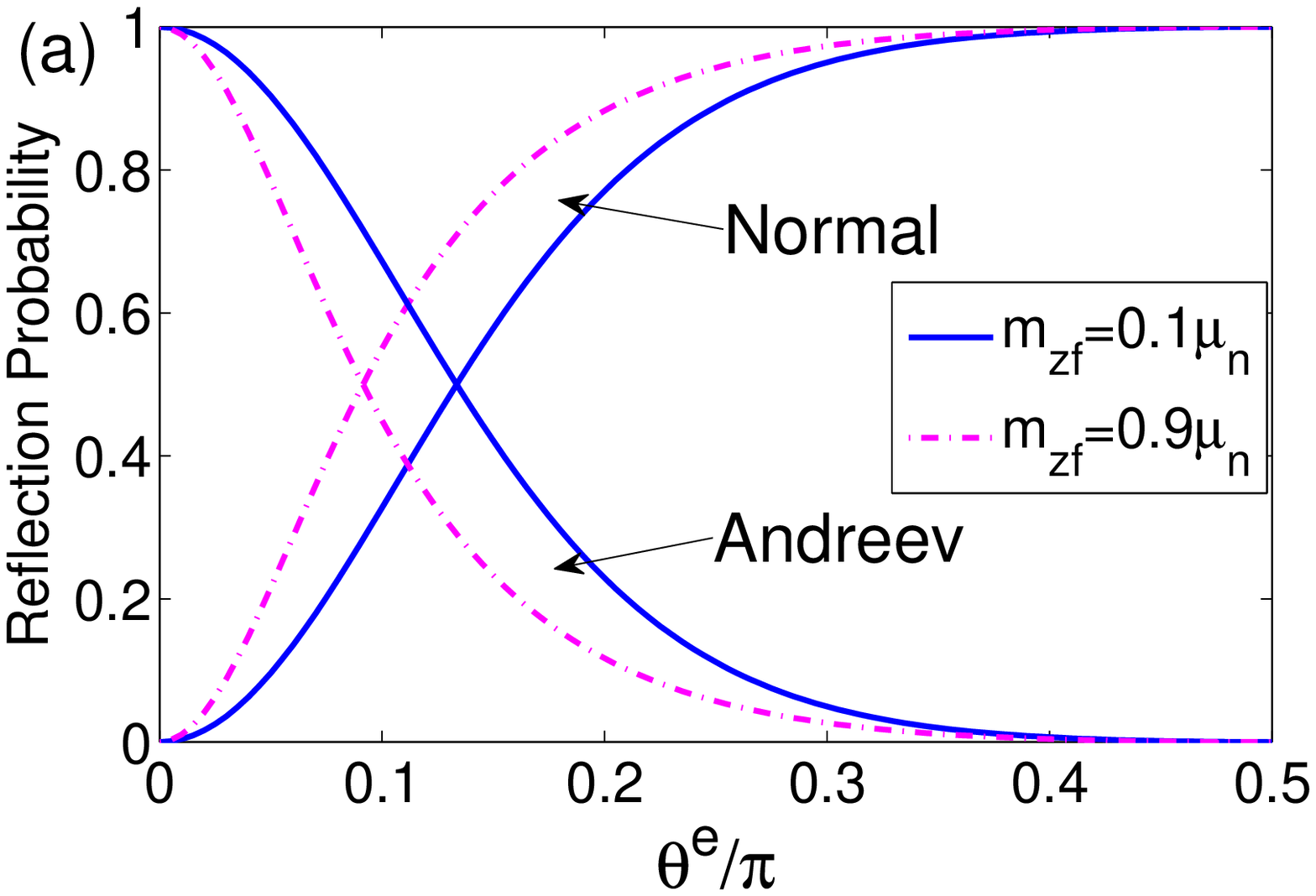}
\end{center}
\end{figure}

\begin{figure}[p]
\epsfxsize=0.5 \textwidth
\begin{center}
\epsfbox{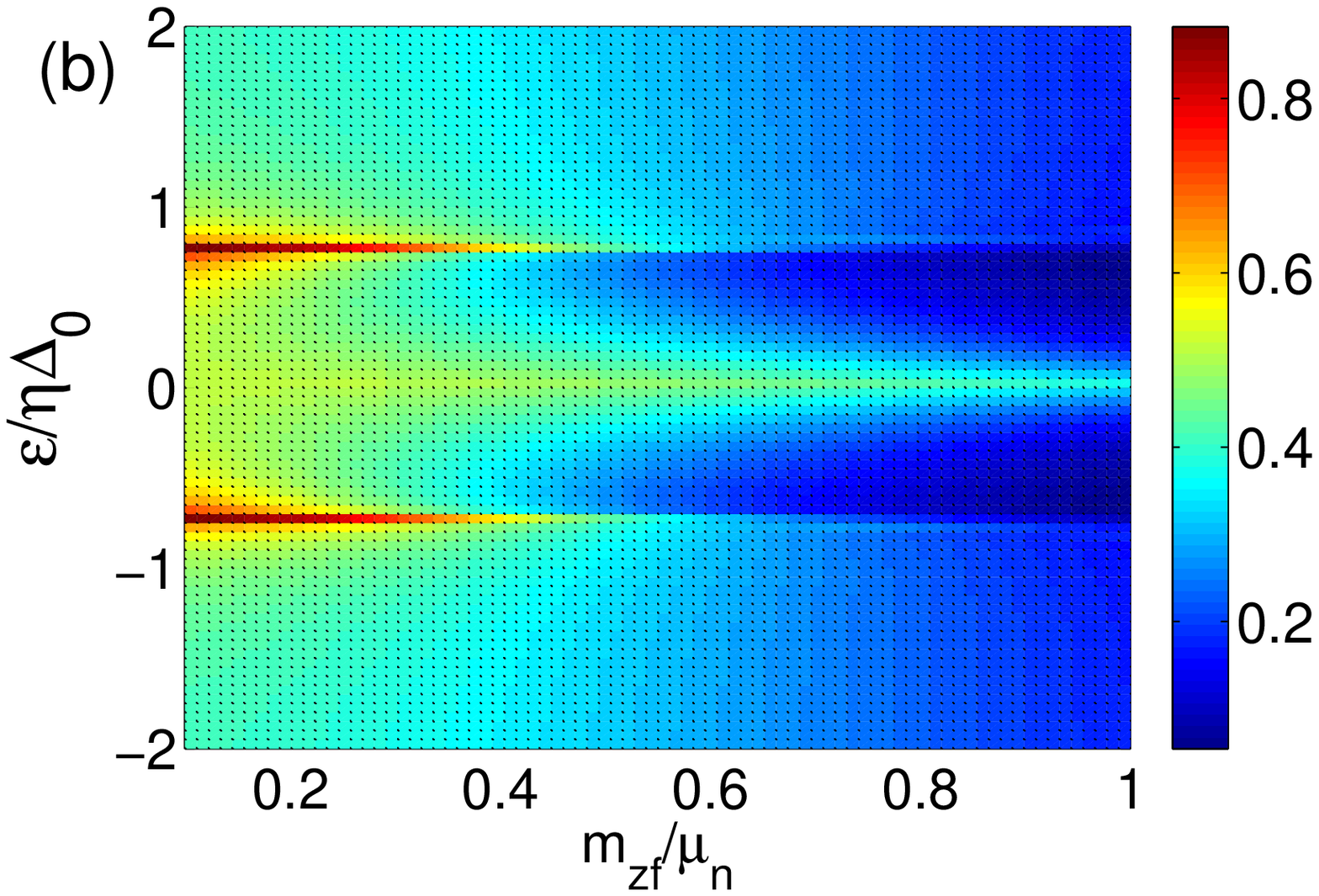}
\end{center}
\end{figure}

\begin{figure}[p]
\epsfxsize=0.5 \textwidth
\begin{center}
\epsfbox{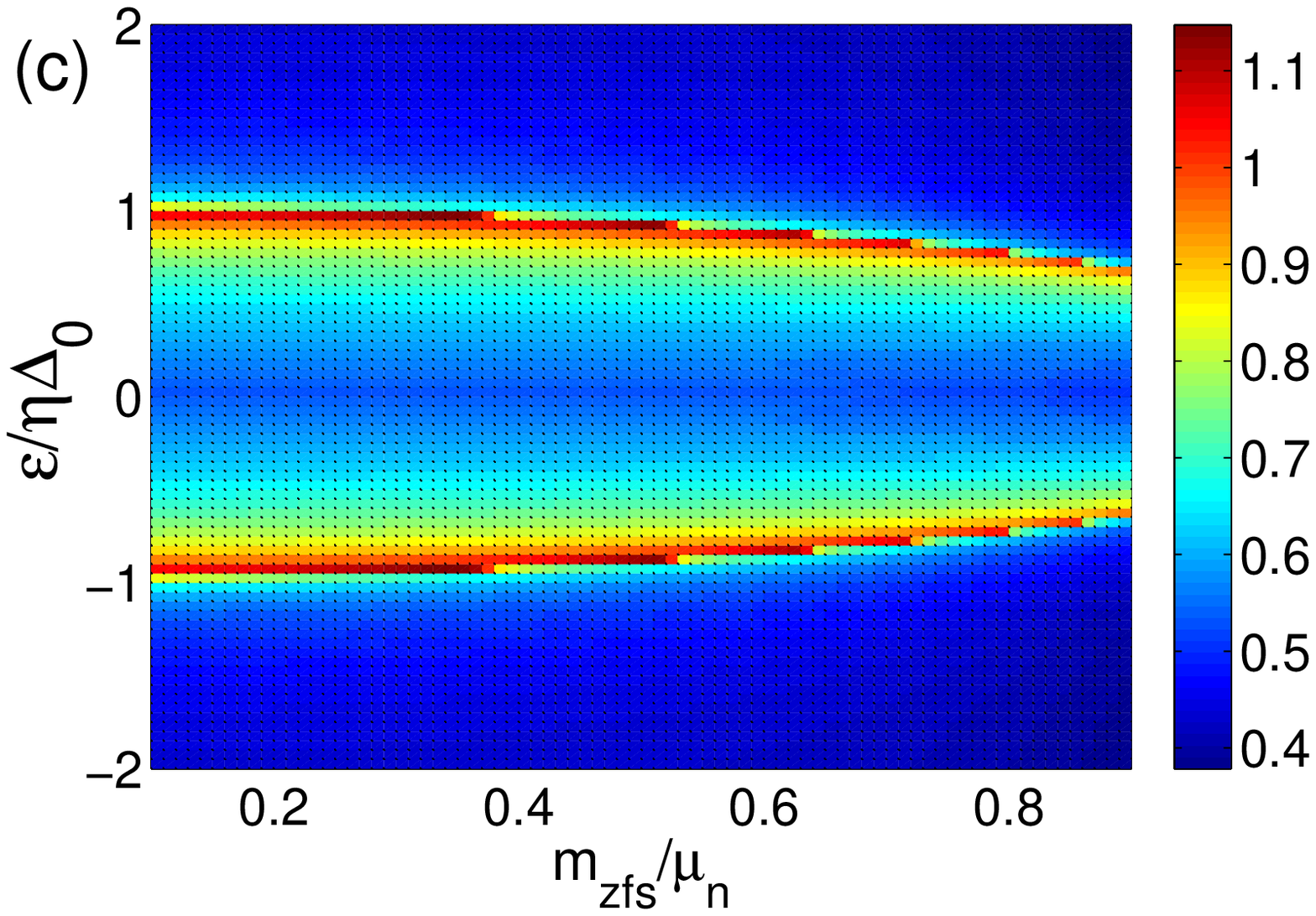}
\end{center}
\end{figure}

\begin{figure}[p]
\epsfxsize=0.5 \textwidth
\begin{center}
\epsfbox{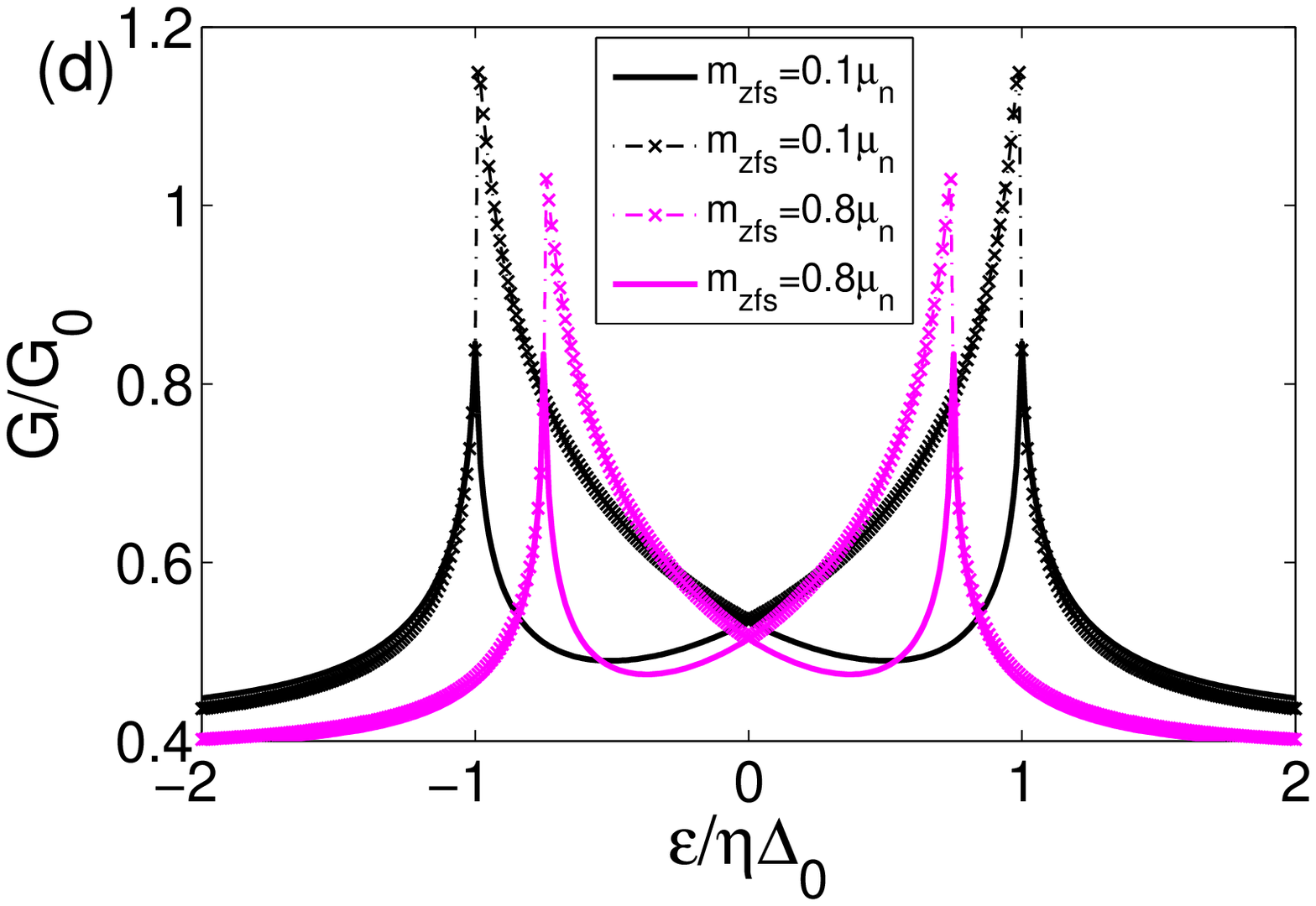}
\setcounter{figure}{3}
\caption{\footnotesize (a),(b),(c),(d)}
\end{center}
\end{figure}

\begin{figure}[p]
\epsfxsize=0.5 \textwidth
\begin{center}
\epsfbox{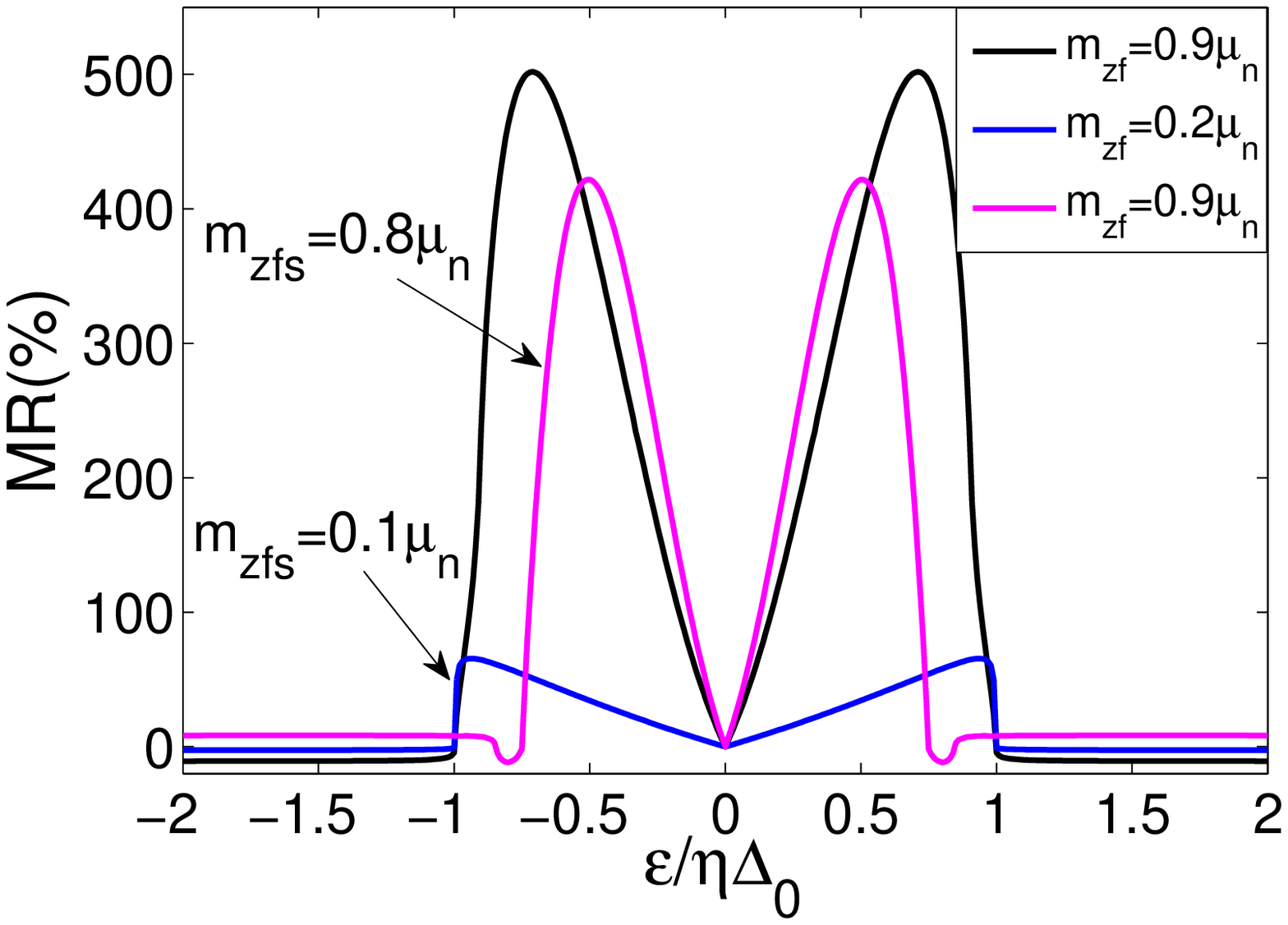}
\setcounter{figure}{4}
\caption{\footnotesize }
\end{center}
\end{figure}

\end{document}